\documentclass[12pt]{JHEP3}
\usepackage{cite}
\usepackage{enumerate}
\usepackage{amsbsy}
\usepackage{graphicx}
\usepackage{rotating}
\usepackage{epsfig}

\newcommand{\be}{\begin{equation}} \newcommand{\ee}{\end{equation}}
\newcommand{\bea}{\begin{eqnarray}} \newcommand{\eea}{\end{eqnarray}}
\newcommand{\el}{\nonumber \\}
\newcommand{\re}[1]{(\ref{#1})}

\renewcommand{\sec}[1]{section \ref{#1}}
\newcommand{\fig}[1]{figure \ref{#1}}
\newcommand{\brt}[1]{[#1]}

\renewcommand{\c}{\gamma}
\renewcommand{\d}{\delta}

\newcommand{\si}{\sigma}
\newcommand{\CC}{{\mathcal C}}

\newcommand{\GN}{G_{\mathrm{N}}}
\newcommand{\ha}{\frac{1}{2}}

\newcommand{\zre}{z_{\mathrm{ri}}}
\newcommand{\zeq}{z_{\mathrm{eq}}}
\newcommand{\rmd}{\mathrm{d}}
\newcommand{\lsim}{\,\raisebox{-0.6ex}{$\buildrel < \over \sim$}\,}
\newcommand{\nonum}{\\}
\newcommand {\etal} {et al.}

\newcommand{\bx}{\boldsymbol{x}}

\newcommand{\bn}{\boldsymbol{n}}
\newcommand{\lmin}{\ell_{\mathrm{min}}}

\newcommand{\La}{\Lambda}
\newcommand{\om}{\omega}
\newcommand{\Om}{\Omega}
\newcommand{\Omm}{\Omega_{\mathrm{m}}}

\newcommand{\omm}{\omega_{\mathrm{m}}}

\newcommand{\ob}{\omega_{{\mathrm{b}}}}
\newcommand{\oc}{\omega_{{\mathrm{c}}}}

\newcommand{\PRD}[1]{{\it Phys. Rev.} {\bf D#1}}

\newcommand{\PRL}[1]{{\it Phys. Rev. Lett.} {\bf #1}}

\newcommand{\PLB}[1]{{\it Phys. Lett.} {\bf B#1}}
\newcommand{\MNRAS}[1]{{\it Mon. Not. Roy. Astron. Soc.} {\bf #1}}
\newcommand{\APJ}[1]{{\it Astrophys. J.} {\bf #1}}
\newcommand{\APJS}[1]{{\it Astrophys. J. Suppl.} {\bf #1}}

\newcommand{\JHEP}[1]{{\it JHEP} {\bf #1}}

\newcommand{\GRG}[1]{{\it Gen. Rel. Grav.} {\bf #1}}
\newcommand{\AaA}[1]{{\it Astron. \& Astrophys.} {\bf #1}}

\title{Model-independent cosmological constraints from the CMB}

\author{Marc Vonlanthen
\\ Universit\'e de Gen\`eve, D\'epartement de Physique Th\'eorique \\
24 quai Ernest-Ansermet, CH-1211 Gen\`eve 4, Switzerland \\
\email{marc {\it dot} vonlanthen {\it at} unige {\it dot} ch}}

\author{Syksy R\"{a}s\"{a}nen
\\ Physics Department, Washington University\\ St Louis, MO 63130, USA
\vspace{0.1cm}
\\ CERN, Physics Department Theory Unit \\ CH-1211 Gen\`eve 4  23, Switzerland
\vspace{0.1cm}
\\ Universit\'e de Gen\`eve, D\'epartement de Physique Th\'eorique \\
24 quai Ernest-Ansermet, CH-1211 Gen\`eve 4, Switzerland \\
\email{syksy {\it dot} rasanen {\it at} iki {\it dot} fi}}

\author{Ruth Durrer 
\\ Universit\'e de Gen\`eve, D\'epartement de Physique Th\'eorique \\
24 quai Ernest-Ansermet, CH-1211 Gen\`eve 4, Switzerland \\
\email{ruth {\it dot} durrer {\it at} unige {\it dot} ch}}

\abstract{
We analyse CMB data in a manner which is as
independent as possible of the model of late-time cosmology. 
We encode the effects of late-time cosmology into a single parameter 
which determines the distance to the last scattering surface.
We exclude low multipoles
$\ell<40$ from the analysis. We consider the WMAP5 and ACBAR data.
We obtain the cosmological parameters $100\ob =2.13\pm 0.05$, $\oc=0.124\pm 
0.007$, $n_s=0.93\pm 0.02$ and $\theta_A=0.593^{\circ}\pm 0.001^{\circ}$
(68\% C.L.). The last number is the angular scale
subtended by the sound horizon at decoupling.
There is a systematic shift in the parameters as more low
$\ell$ data are omitted, towards smaller values of $\ob$ and $n_s$
and larger values of $\oc$. The scale $\theta_A$ remains stable
and very well determined.
}

\preprint{CERN-PH-TH/2010-053}
\begin{document}
\setcounter{tocdepth}{2}
\setcounter{secnumdepth}{3}

\section{Introduction} \label{sec:intro}

The cosmic microwave background (CMB) is one of the most
important cosmological probes.
The pattern of acoustic oscillations of the baryon-photon
plasma is imprinted on the CMB at the time of decoupling,
and then rescaled (and on large scales modified) as the 
CMB photons propagate from the last scattering surface
to the observer. The CMB is thus sensitive to cosmological
parameters in two ways, via the physics at decoupling
and via the evolution of the universe after that.

While the physics at decoupling --essentially atomic physics
and general relativity of a linearly perturbed
Friedmann-Lema\^\i tre (FL) universe-- is well understood,
the evolution at late times deviates from the predictions
of linearly perturbed FL models with radiation and matter.
The difference may be due to an exotic matter component with
negative pressure such as vacuum energy, deviation of gravity 
from general relativity~\cite{RR,Linder,Koyama,CapoFranca},
or a breakdown of the homogeneous and isotropic approximation
\cite{Rasanen:2006b,Buchert:2007,Enqvist:2007,Rasanen:2008a,Zibin:2008b,Celerier:2009}.
It is not known which of these possibilities is correct,
and there are large differences between the various models.
It is therefore worthwhile to analyse the CMB in a manner
which is as independent of the details of late-time cosmology
as possible.
On the one hand, this clarifies the minimal constraints
that all models of late-time cosmology, whatever their details,
have to satisfy in order to agree with CMB observations.
On the other hand, our analysis provides limits on the
physical parameters at decoupling that are independent of
the details of what happens at later times.
This is particularly important for cosmological parameters
such as the density of baryons, density of dark matter and
the spectral index, which are used to constrain particle physics
models of baryogenesis, supersymmetry and inflation, which are
independent of late-time cosmology.

Such a separation of constraints is possible because
the physics after decoupling affects the CMB in a
rather limited manner (except at low multipoles),
by simply changing the angular scale and modifying
the overall amplitude of the CMB pattern.
We encode the change in the angular scale in a single
parameter related to the angular diameter distance to
the last scattering surface and treat the
amplitude as a nuisance parameter.
We aim to be transparent about how the different
cosmological parameters enter the calculation and
the assumptions that go into the analysis.

In section 2 we discuss how the physics at early
and late times affects the CMB and explain our
assumptions. In section 3 we present the results
of the analysis of the WMAP 5-year data
\cite{Komatsu:2008, Dunkley:2008, Hinshaw:2008}
and the ACBAR data \cite{acbar} and give
the constraints on cosmological parameters.
In section 4 we summarise our results.
Some details are collected in two appendices.

\section{Parameter dependence of the CMB}

\subsection{Our assumptions}

The pattern of CMB anisotropies can be summarised
in terms of a few parameters.
It was noted in \cite{Efstathiou:1998} that
models with the same primordial perturbation spectra
and same values of $\ob, \oc$ and the shift parameter
$R$ have an identical CMB spectrum today, apart from
low multipoles ($\ell\lesssim30$).
The discussion in \cite{Efstathiou:1998} was in the context
of a family of Friedmann-Lema\^\i tre (FL) models,
but the statement is true more generally.
The shift parameter is defined as
\bea \label{R}
  R &\equiv& \omm^{1/2} (1+z_*) H_0 h^{-1} D_A(z_*) \el
  &=& \left(\frac{\Om_m}{\Omega_{K}}\right)^{1/2} \sinh \left(\Omega_{K}^{1/2} \int_0^{z_*} \rmd z' \frac{H_0}{H(z')} \right) \ ,
\eea

\noindent where $z_*$ is the redshift of decoupling,
$D_A(z)$ is the angular diameter distance between today
and redshift $z$, $H_0=100 h$km/s/Mpc
is the Hubble parameter today, and the second equation holds for
all FL models.
The density parameter $\ob$ is the normalized dimensionless
physical density of baryonic matter,
$\ob=8\pi\GN\rho_b/3$/(100 km/s/Mpc)$^2$,
$\oc$ is the normalized dimensionless physical density of cold
dark matter defined the same way, $\omm=\ob+\oc$ is the total
physical matter density, and $\Omm=\omm h^{-2}$ and
$\Omega_{K}$ are, respectively,
the matter and the spatial curvature density parameter today.
With present observations which include polarization
data, one has to add a parameter to take into account collisions
between the CMB photons and baryonic matter after the cosmic
medium becomes reionized. This is usually expressed with the
redshift of reionization $\zre$ or the optical depth $\tau$.

The CMB data have been analysed in terms of the shift parameter 
$R$ in various FL models
\cite{Kosowsky:2002, Komatsu:2008, Elgaroy:2007, Wang:2007, Corasaniti:2007, Mukherjee:2008, Komatsu:2010},
and a similar approach has been followed for local void models
\cite{Zibin:2008b, Clifton:2009}.
The model-dependence of parameters such as $R$ has been discussed,
but limits on them have always been derived within some
specific models, and it has not been clear which assumptions are
important and what is the model-independent information.

In this work, we analyse the CMB in a manner which is as model-independent
as possible, and we are explicit about the assumptions involved.
In particular, we do not restrict our study to models which are close
to FL at late times, so our constraints are also applicable to
models where the effect of non-linear structures on the expansion
rate is important, or where we are located in a large spherically
symmetric density fluctuation such as a local void.
(Note that the near-isotropy
of the CMB does not imply that the universe is close to FL,
even coupled with the Copernican principle \cite{Rasanen:2009}.)

We assume that the physics up to and including decoupling
is completely standard, i.e. linearly perturbed FL evolution
according to normal four-dimensional general relativity with
Standard Model particle physics and dark matter
(which we assume to be cold during decoupling).
As for physics after decoupling, we make the minimal assumptions
that it changes the small angle CMB spectrum only by 1) modifying
the angular diameter distance to the last scattering surface and
2) changing the overall amplitude.
Here, small angles refers to scales which are well
inside the horizon at late times when the unknown physics
can be important, say conservatively at $z\lsim 60$. 
We discard low multipoles in our analysis, because
typically the unknown physics of dark energy, modified
gravity or large deviations from FL geometry affects the
large angles in a model-dependent way, for example via the
late Integrated Sachs-Wolfe (ISW) effect.
In typical perturbed FL models, the late ISW effect is only
significant at low multipoles (see appendix \ref{ap:scale}),
and the Rees-Sciama effect, gravitational lensing and the
Sunyaev-Zel'dovich effect do not have a significant
impact at the present observational accuracy
\cite{Lewis:2006, Durrer:2008, Seljak:1996}, though
their presence is already suggested by the ACBAR data \cite{acbar}.
We assume that such effects remain small in other models,
and that any multipole-dependent effect of new physics
on the CMB spectrum is below
the observational precision, except at low multipoles.

In perturbed FL models, reionization has a significant
effect on all angular scales, but at high multipoles it
amounts to a simple rescaling of the amplitude, and
is thus degenerate with the amplitude of primordial
perturbations (see appendix~\ref{ap:reion}), so we can neglect
modeling of reionization. 

We assume that the primordial
perturbations are adiabatic, and have a power-law spectrum.
We only consider scalar perturbations, and assume that 
vector and tensor contributions are small. (This division
refers to the early universe; in the late universe it is not
necessarily meaningful, because we do not assume that the
late universe is close to FL.)
Within our approach it would not be easy to include tensor 
perturbations in the temperature anisotropy spectrum, because
they contribute  mainly via the ISW effect and are relevant up to 
$\ell\approx100$. However, the contribution of tensors starts
to decay already around $\ell\approx 50$ and is probably relatively
small, so their presence would not be expected to change our results
significantly. (It would be easy to take into account the tensor
contribution to the polarization spectrum, though, because it is mainly
generated at the last scattering surface.)
We also neglect the effect of neutrino masses.

The idea behind these assumptions is that we can treat
the CMB with a standard Boltzmann code, and simply exclude
low multipoles from the analysis.
We have modified the publicly available CAMB code
and the corresponding Monte Carlo Markov Chain 
program~\cite{CAMB} to search for best-fit values of
our parameters.
As long as the rise to the first peak is fully included in the analysis,
discarding low multipoles should not involve a significant loss of 
information, because there are more high multipoles and the
cosmic variance is larger on large scales.
However, our results in this respect are somewhat
surprising, as we discuss in \sec{sec:results}.
Also, it has been argued that there are anomalies in
the angular distribution on large scales \cite{anom}
(and a dipolar modulation at higher multipoles \cite{Hansen:2008}),
which could indicate that some physics affecting the low
multipoles is not understood, so they may be
unreliable for cosmological analysis;
see also \cite{Francis:2009, Bennett:2010}.

Our assumptions do not hold for models with non-standard physics
at or before decoupling, such as new radiation degrees of freedom,
early dark energy \cite{earlyDE} or dark matter which undergoes
significant annihilation at early times \cite{DMann}.
In models where we are located in a large spherically
symmetric region, it is possible to obtain a large CMB
dipole \cite{LTBdipole}, and there could be a large effect at higher
multipoles as well. This can only be checked with perturbation
theory adapted to such models, which is now being developed \cite{LTBpert}.

\subsection{The physics of the CMB parameters}

Let us outline the relation between the
features in the CMB spectrum and the cosmological
parameters, given our assumptions above.
(See \cite{Durrer:2008, Mukhanov} for detailed discussion.)
We consider five parameters, namely the overall amplitude,
the baryon density $\ob$, the cold dark matter density $\oc$,
the spectral index $n_s$ and the distance to the last
scattering surface $D_A(z_*)$.

The observed amplitude of CMB perturbations is
determined by a combination of the primordial power
spectrum and late-time physics, such as damping
due to accelerating expansion and scattering
of CMB photons from matter due to reionization. Without specifying
a model for the late-time universe, it is not
possible to disentangle these effects.
Because the overall normalization does not have
a model-independent interpretation, we treat it as a nuisance parameter, 
i.e., we marginalize over it and do not quote limits for it.

The spectral index $n_s$ is related to the early universe
physics, such as inflation, which produces the primordial
perturbations. Extending the analysis to more complicated
primordial spectra would be straightforward, though of course
we would not be sensitive to large-scale features.

The relative height and depth of the CMB peaks
and troughs is set by the physics of the
baryon-photon oscillations, which depends
on $\ob$ and $\oc$.
This pattern also depends on the radiation density
$\omega_r=8\pi\GN\rho_r/3$/(100 km/s/Mpc)$^2$,
which is however accurately determined by the CMB temperature.
Note that the CMB is only sensitive to the densities
at the time of decoupling, not to their values today.
As is customary, we use the symbols $\ob$, $\oc$ and $\omega_r$
to refer to the densities at decoupling scaled
to today with the factor $(1+z_*)^3$ for baryons
and dark matter and $(1+z_*)^4$ for radiation,
where $*$ indicates the time of decoupling.
At decoupling, the distribution of
matter is still very smooth, so the densities
at that time can be understood as local or
average values; the scaled numbers
represent today's average values.
In a statistically homogeneous and isotropic space,
the mean energy density of baryons and cold dark matter
evolves like $(1+z)^3$ due to conservation of mass,
and the mean energy density of photons evolves like
$(1+z)^4$ due to conservation of photon number
and the fact that the change of energy of the CMB
photons by scattering can be neglected \cite{Rasanen:2008b}.
FL models are of course a particular case of this.
If dark matter has significant pressure, or decays
significantly \cite{DMdecay}, or if there is some
extra source of baryons, dark matter or photons,
our $\ob, \oc$ and $\omega_r$ would not correspond to the
physical densities today.
(Dark matter decay to radiation would also contribute
to the late ISW effect \cite{DMdecayISW}.)
This is already true for neutrinos, which we treat as
massless, but which in fact do not contribute
to the present-day radiation density, since their
mass today is larger than the temperature.
This will also be the case if the factor $(1+z)^3$
is not simply proportional to the volume, which can
happen if statistical homogeneity and isotropy is
broken, such as in local void models where shear
can contribute significantly to the redshift.

Our final parameter is the angular diameter distance
to the last scattering surface. The
angular diameter distance out to redshift $z$ is defined
as $D_A(z)=L/\theta$, where $L$ is the proper size of
an object at redshift $z$ and $\theta$ is its observed
angular size. The physical scale of the baryon-photon
oscillations is set by the sound horizon at decoupling
$r_s(z_*)$ which depends on $\ob$ and $\oc$ \cite{Hu:1995, Durrer:2008}.
With standard physics up to decoupling, the sound speed of
the photon-baryon plasma is
\bea
  c_s^2 = \frac{1}{3(1+3\rho_b/4\rho_\c)} = \frac{1}{3\left(1+\frac{3\ob}{(1+z) 4
\omega_\c}\right)} \equiv \frac{1}{3[1+r (1+z)^{-1}]} \ ,
\eea

\noindent where we have introduced $r\equiv 3\ob/4\omega_\c$.
For the sound horizon we obtain\footnote{Here $r_s$ is the physical
sound horizon at the time of decoupling. In the literature, $r_s$
often denotes the comoving sound horizon.}
\bea
  (1+z_*)r_s(z_*) &=& \int_0^{t_*}\frac{c_s(t')}{a(t')}dt' \el
  &=& \frac{h}{H_0\sqrt{3}} \int_{1+z_*}^\infty\frac{dx}{x\sqrt{(x+r)
    (x\omega_r +\omm )}} \nonumber \\ 
  &=& \label{rsz*}  \frac{2h}{H_0\sqrt{3r\omm}}
\log\left(\frac{\sqrt{1+z_* +r} +\sqrt{\frac{(1+z_*)
r\omega_r}{\omega_m} +r}}{\sqrt{1+z_*}\left(1 +
 \sqrt{\frac{r\omega_r}{\omm}}\right)}\right) \ .
\eea

\noindent Note that $h/H_0=1/(100$km/s/Mpc$) \approx 2998$ Mpc
is a fixed scale which does not depend on the cosmological model.
The photon energy density $\omega_\c\approx 2.48\times 10^{-5}$
is known as well as the CMB temperature and we do not treat
it as a free  parameter. Assuming massless neutrinos, the
same is true for the radiation density \cite{Durrer:2008},
$\omega_r = \omega_\gamma\left(1+3\frac{7}{8} \left(\frac{4}{11}\right)^{4/3}\right) 
\approx 4.17\times 10^{-5}.$
Furthermore, for standard radiation content, $z_*\approx 1090$ and it depends
weakly on $\ob$ and $\oc$ (for an  analytical approximation,
see \cite{Hu:1995}). For standard values of the parameters, the log 
in (\ref{rsz*}) is of order unity.
The sound horizon at decoupling therefore depends only on
$\ob$ and $\oc$. The angle under which it is
observed today is given by $\theta_A\equiv r_s(z_*)/D_A(z_*)$.
With $\ob$ and $\oc$ fixed, the pattern of CMB anisotropies
is determined at decoupling (apart from low multipoles), and
its angular scale changes as the distance to the last scattering
surface grows and the multipole positions of the
CMB peaks and troughs scale with $D_A(z_*)$.
Given our assumptions, the CMB (apart from low multipoles) 
has no sensitivity to any physical parameters other than
$\ob, \oc, n_s, D_A$ and the overall amplitude, and
these five parameters are a priori independent.
A given model can of course couple them to each other,
as well as to parameters which do not directly affect the CMB.

In particular, in linearly perturbed FL models
the spatial curvature affects the CMB only via the angular
diameter distance (apart from the late ISW effect).
It is sometimes said that the spatial curvature can be
determined from CMB observations by using the sound horizon
as a standard ruler (assuming that the universe can be
described by a FL model). However, as \re{R} shows,
the effect of spatial curvature on $D_A(z_*)$ is completely
degenerate with the expansion history $H(z)$.
For example, FL models with matter and significant
spatial curvature are consistent with the WMAP
observations \cite{Spergel:2006}. In that case, a prior on $H_0$
is enough to exclude large spatial curvature, but only
because of the specific form of the expansion history.
The only way to really measure the spatial curvature, as
opposed to doing parameter estimation in the context of
specific models, is to use independent observations of the 
distance and expansion rate~\cite{chris}, such as from
the ages of passively evolving galaxies~\cite{ages}
and baryon acoustic oscillations~\cite{BAO}.
Note also that the CMB (apart from low multipoles) is
sensitive to the expansion history between
decoupling and today only via the angular diameter distance;
in particular, the CMB contains no model-independent
information about $H_0$.

In addition to $R$, another parameter defined as
\bea \label{la}
  \ell_A \equiv \frac{\pi}{\theta_A} = \pi \frac{D_A(z_*)}{r_s(z_*)}
\eea
has also been introduced to parametrise the
distance to the last scattering surface \cite{Hu:2000}.
The parameter $\ell_A$ is related to the position of
the first peak in multipole space
(for details, see \cite{Durrer:2008, Mukhanov, Hu:2000, Doran:2001}).
The quantity $\ell_A$ has been called an independent
shift parameter in addition to $R$ \cite{Wang:2007}.
However, this is somewhat misleading, because $R$ and $\ell_A$
contain the same information as regards the shift in the
angular scale of the CMB anisotropy pattern due to the
late-time evolution, the only difference is their
dependence on $\ob$ and $\oc$.
Of course, one can consider any combination of the
four parameters $\ob, \oc, n_s$ and $R$.
For our purposes, it is useful to introduce the scale
parameter $S$, which is defined as the ratio of the
angular diameter distance to the prediction of the
simplest cosmological model,
\bea \label{S}
  S &\equiv& \frac{D_A(z_*)}{D_{A,EdS}(z_*)} = \frac{ H_0 (1+z_*) D_A(z_*) }{2[1-(1+z_*)^{-1/2}]} \simeq \ha H_0 (1+z_*) D_A(z_*) \ ,
\eea

\noindent where $D_{A,EdS}$ is the angular diameter distance
in the Einstein-de Sitter (EdS) universe (the matter-dominated
spatially flat FL model), $(1+z) D_{A,EdS}=2 H_0^{-1} [1-(1+z)^{-1/2}]$;
the last approximation in \re{S} is accurate to 3\%.
Using \re{R}, the scale parameter $S$ is related to $R$ by
$S=h R/(2\omm^{1/2}) [1-(1+z_*)^{-1/2}]^{-1}\simeq h R/(2\omm^{1/2})$.
Unlike $R$ and $\ell_A$, the scale parameter $S$ depends on
the Hubble parameter, to which the CMB has no direct
sensitivity. (This arises because FL models predict the
distance in units of $H_0$.)
Therefore, the value of $S$ depends on how we fix the
Hubble parameter.

We can simply keep $H_0$ free and quote
limits for $h^{-1} S$, and one can then substitute the
Hubble parameter given by e.g. local observations of $H_0$.
The mean value is $h^{-1} S=2.4$ (see table \ref{t:res}),
so for $h=$ 0.6--0.7, the distance to  the last scattering
surface is a factor of 1.4--1.7 longer than in an EdS model
with the observed Hubble parameter.
This is in accordance with the usual intuition
that physics in the late-time universe acts to
increase the distance compared to EdS, for example
via accelerated expansion.
We could instead keep the age of the universe fixed,
i.e. ask how large the distance is compared to the
value in an EdS model at the same time after the big bang.
In an EdS model $H_0=2/(3 t_0)$, so we have
$S=2/(3 t_0 100\mathrm{km/s/Mpc}) \times R/(2\omm^{1/2}) [1-(1+z_*)^{-1/2}]^{-1}$,
which for $t_0=13.4$ Gyr \cite{Krauss:2003}
gives $S\approx 1.2$ for our mean values $\omm=0.145$ and $R=1.77$.

Finally, we can ask how long the distance is compared to
an EdS model which has the correct matter density.
The Hubble parameter is then simply $h=\omm^{1/2}$,
so $S=R/2 [1-(1+z_*)^{-1/2}]^{-1}\simeq R/2 \approx 0.9$.
This means that in an EdS model with the correct matter density, the 
predicted distance to the last scattering surface is {\it longer} than
observed. (In other words, the real matter density decays faster as 
function of the distance to the last scattering surface than in the 
EdS reference model.) Unless otherwise noted, we follow this last convention,
and compare with an EdS model which has the correct matter density, at the 
expense of the age of the universe and the Hubble parameter.
We give constraints for
$\theta_A, \ell_A, R$, $S$, $h^{-1}S$,  and $D_A(z_*)$ in~\sec{sec:results}.
For fixed $\ob$, $\oc$ and $n_s$, these quantities contain the same 
information, only their correlation properties with the parameters
 $\ob$, $\oc$ and $n_s$ are different (see table \ref{t:cov}).

\subsection{The distance to the last scattering surface} \label{sec:method}

Let us now study how the CMB spectrum depends on the angular diameter 
distance to the last scattering surface $D_A(z_*)$. We consider two 
positions on the sky denoted by $\bn_1$ and $\bn_2$ which have the
temperature fluctuations $\Delta T(\bn_1)$ and $\Delta T(\bn_2)$ and which 
are separated by proper distance $L$ on the last scattering surface.
For two different angular diameter distances $D_A$ and $D_A'$ to the last 
scattering surface, the length $L$ is seen under the angles 
$\theta=L/D_A$ and $\theta'=L/D'_A$, see~\fig{f:DA}.

\FIGURE[ht]{
\hspace*{2cm} \epsfig{width=8cm, file=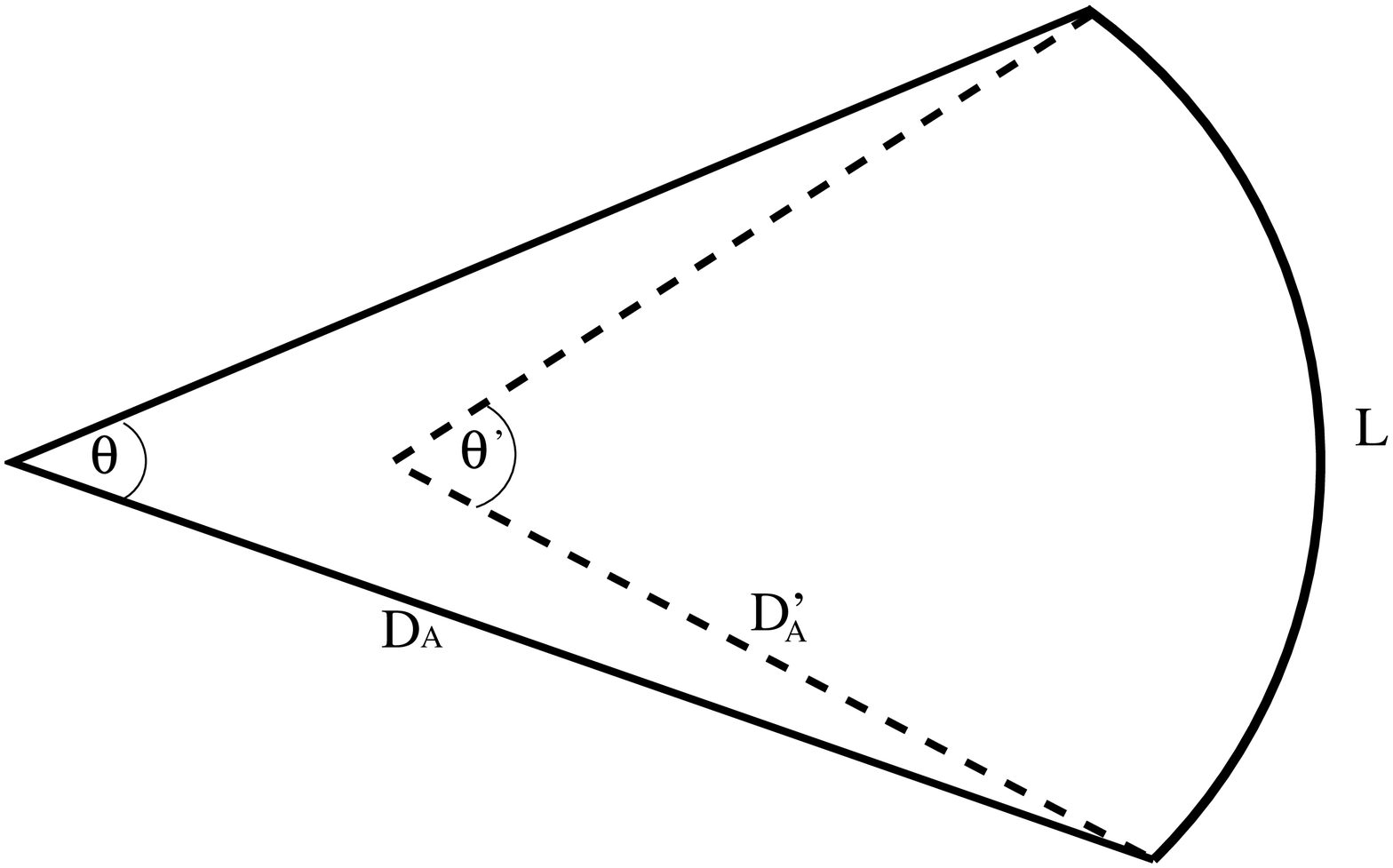} \hspace*{2cm}
\caption{The angle under which two fixed points on the sky are seen 
changes with the angular diameter distance $D_A$.
 }\label{f:DA}}

The two-point functions $\CC$ and $\CC'$ which correlate
$\bn_1$ and $\bn_2$ for an observer at distance $D_A$ or $D_A'$,
respectively, can be decomposed in terms of the two angles as
\bea
  \CC(\theta) \equiv \langle\Delta T(\bn_1)\Delta T(\bn_2)\rangle
&=& \frac{1}{4\pi}\sum_{\ell}(2\ell+1)C_\ell P_\ell(\cos\theta)
\nonumber \\
&=&  \frac{1}{4\pi}\sum_{\ell}(2\ell+1)C'_\ell P_\ell(\cos\theta')
 =  \CC'(\theta') \,,
\eea

\noindent where $P_\ell$ is the Legendre polynomial of degree
$\ell$, and $C_\ell$ and $C'_\ell$ are the power spectra
corresponding to the angular diameter distances $D_A$ and $D_A'$
respectively. The equality $ \CC(\theta) = \CC'(\theta')$ means that we 
consider only correlations on the last scattering surface (or very close to it)
and neglect line-of-sight effects like, e.g. the late ISW effect which can be 
different for the two photon paths. Using the orthogonality of 
the Legendre polynomials,
$\int_{-1}^1P_{\tilde\ell}(\mu)P_\ell(\mu)d\mu =2 \d_{\ell,\tilde\ell}/(2\ell+1)$,
we obtain the relation
\be \label{e:clcl'gen}
  C_\ell = \sum_{\tilde\ell} \frac{2\tilde\ell+1}{2} C'_{\tilde\ell}\int_{0}^\pi
\sin\theta d\theta P_{\tilde\ell}[\cos(\theta D_A/D_A')] P_\ell(\cos\theta)
\ .
\ee
\noindent This cumbersome exact expression is only needed for
low values of $\ell$. At high $\ell$ we can work in the flat sky 
approximation (see~\cite{Durrer:2008}, section 5.4), where
$$ Y_{\ell m} \rightarrow \frac{1}{2\pi}\exp(i{\boldsymbol{\ell}}\cdot\bx) \quad
\mbox{ and } P_\ell(\cos\theta) \rightarrow J_0(|\bx|\ell) \ . $$
Here $\bx$ is a vector on the flat sky,  $\boldsymbol{\ell}$ is the
variable of its 2-dimensional Fourier transform, with
$\ell=|\boldsymbol{\ell}|$, and $J_0$ is the Bessel function
of order $0$. Denoting $r\equiv|\bx|$, the correlation function is
\be
  \CC(\theta) = \CC(r) = \frac{1}{2\pi}\int_0^\infty \! \! d\ell\, 
   \ell J_0(r\ell)C_\ell \ .
\ee
\noindent The correlation functions corresponding to the two
angular diameter distances are related by $\CC(r)=\CC'(r')$,
where $r'=rD_A/D'_A$,
\bea
  \frac{1}{2\pi}\int_0^\infty \!\! d\ell\,\ell J_0(r\ell)C_\ell &=& 
\frac{1}{2\pi}\int_0^\infty\!\! d\ell\,\ell  J_0(r'\ell)C'_\ell \nonumber\\
  &=& \frac{1}{2\pi}\left(\frac{D'_A}{D_A}\right)^2\int_0^\infty \!\! d\ell\, 
      \ell J_0(r\ell) C'_{\frac{ D_A'}{D_A}\ell} \ , \label{e:clcl'}
\eea

\noindent where on the second line we have simply performed the change
of variables $\ell\rightarrow \ell D_A/D'_A$. Using the property
$\int_0^\infty rdr J_0(r\ell)J_0(r\ell') = \ell^{-1}\delta(\ell-\ell')$,
we obtain
\bea \label{Clrel}
  C_\ell &=&  \left(\frac{D_A'}{D_A}\right)^2 C'_{\frac{ D_A'}{D_A}\ell}  \ .
\eea

The relation \re{Clrel} is valid independent of spatial
curvature, since we do not invoke three-dimensional Fourier transforms.
We are simply using the fact that the CMB anisotropies are functions on a 
sphere. This result agrees with~\cite{Zibin} where it is derived in a 
different way and contrasts with \cite{Clifton:2009}, where there is an extra 
power of $D_A'/D_A$. 
Let us denote the spectrum of a reference EdS Universe
by $C_\ell^{EdS}$ and the measured CMB spectrum by $C_\ell$.
Recalling the definition \re{S} of the scale parameter $S$,
we can assign $C_\ell$ to an EdS universe with
the same values of $\ob$, $\oc$ and $n_s$ and
the angular diameter distance $D_A = S D_{A,EdS}$
if we scale the angular power spectrum by
\be \label{e:ClS}
  C_\ell = S^{-2}C_{S^{-1}\ell}^{EdS} \ .
\ee 
The basic assumption here is that the CMB fluctuations at 
decoupling are the same for both models and the only difference is 
the distance to the last scattering surface. If this is true,
the relation (\ref{e:ClS}) is exact in the flat sky limit. Without
the flat sky approximation it has to be replaced by
(\ref{e:clcl'gen}) with $D_A/D_A'=S$.
Note that despite of the factor $S^{-2}$ in (\ref{e:ClS}), the shift parameter
$S$ is not strongly correlated with the amplitude, it just {\em shifts} the 
spectrum in angular space. This is visible on the 2D-plots shown 
in Fig.~\ref{f:shift2d}.
We have tested the flat sky approximation numerically and have found that 
for $\ell\geq20$ the difference between (\ref{e:ClS}) and the exact
expression (\ref{e:clcl'gen}) is smaller than 1\% for $1.1\geq S\geq0.7$,
which includes the region which is of interest to us
(the mean value we obtain is $S=0.91\pm0.01$, see table \ref{t:res}).

To illustrate the dependence of the CMB spectra
on the scale parameter $S$, we show in
appendix \ref{ap:scale} the TT, TE and EE spectra
for FL models with non-zero spatial curvature
or cosmological constant, compared with the EdS
result scaled with $S$.
As shown in figures \ref{f:curvatureTT} to \ref{f:curvatureEE},
the spectra for the scaled model and the model with
spatial curvature lie on top of each other for
$\ell\gtrsim 20$, except for the case of large
negative spatial curvature with $S\approx 1.5$, where there is some
difference in the TT spectra until $\ell\approx100$.
For the cosmological constant case, shown in
\fig{f:lambda}, the approximation is excellent for
all of the spectra for $\ell\gtrsim 20$.

\section{Results} \label{sec:results}

\subsection{Cosmological parameters and the multipole cut}

We use the WMAP5 data and the ACBAR data in our analysis.
However, disregarding ACBAR does not change the results much. 
We have performed a Markov Chain Monte Carlo analysis with chain length
$N=2\times 10^5$. The results change by significantly less than $1\si$ 
when going from $N=1.5\times 10^5$ to $2\times 10^5$,
which indicates that the chains have converged well~\cite{CAMB}.
As a convergence test, we have checked that when the samples
are split in two or three parts, the change of the relevant
cosmological parameters is a few percent of one
standard deviation. We have also checked that the
Raftery and Lewis convergence diagnostic is satisfied \cite{rafterylewis1992}.

\TABLE[ht]{
\begin{tabular}{||l|c|c|c|c||}
\hline
$\lmin$ &  2 & 20 & 40 & 60 \\  \hline
$100 \ob$ & $2.21^{+0.05}_{-0.05}$ & $2.19^{+0.05}_{-0.05}$ & 
$2.18^{+0.07}_{-0.07}$ & $2.15^{+0.08}_{-0.08}$ \\
$\oc$ & $0.113^{+0.005}_{-0.005}$ & $0.115^{+0.006}_{-0.006}$ &
$0.118^{+0.007}_{-0.007}$ & $0.120^{+0.008}_{-0.008}$ \\
$n_s$ & $0.95^{+0.01}_{-0.01}$ & $0.95^{+0.01}_{-0.01}$ & $0.94^{+0.02}_{-0.02}$ & 
$0.93^{+0.02}_{-0.02}$ \\
$\Om_{\La}$ & $0.72^{+0.03}_{-0.03}$ & $0.71^{+0.04}_{-0.03}$ & 
$0.70^{+0.04}_{-0.04}$ & $ 0.68^{+0.06}_{-0.05}$ \\
\hline
$\lmin$ & 80 & 100 & 120 & 140  \\ \hline
$100 \ob$ & $2.09^{+0.10}_{-0.10}$ & $2.05^{+0.09}_{-0.09}$ & $2.11^{+0.13}_{-0.12}$ &  $2.07^{+0.14}_{-0.14}$  \\
$\oc$  & $0.127^{+0.012}_{-0.013}$ & $0.132^{+0.012}_{-0.012}$ & $0.126^{+0.013}_{-0.016}$ & $0.131^{+0.018}_{-0.017}$  \\
$n_s$ & $0.91^{+0.03}_{-0.04}$ & $0.89^{+0.04}_{-0.03}$ & $0.91^{+0.05}_{-0.04}$ & $0.90^{+0.05}_{-0.06}$ \\
$\Om_{\La}$ & $0.62^{+0.09}_{-0.09}$ & $0.58^{+0.10}_{-0.09}$ & $0.63^{+0.11}_{-0.10}$ & $0.58^{+0.14}_{-0.14}$  \\
\hline
\end{tabular} 
\caption{\label{t:linfo} The change in the mean parameters 
when more low $\ell$ data are omitted, in the $\Lambda$CDM model
with $\tau=0$. We have used the WMAP5 and ACBAR data.}
}

In table \ref{t:linfo} we show the effect of excluding
a successively larger multipole range up to $\lmin$
in the analysis of the $\Lambda$CDM model; $\Omega_\La$
is the vacuum energy density today, as usual.
We have set $\tau=0$ for consistency with the treatment
of the scaled model.
From $\lmin=2$ to $\lmin=40$ the errors on $\ob$ and
$\oc$ increase by 28\%, while the error on $n_s$ increases by 57\%.
The central values move only by 1\%, 4\% and 1\%, respectively,
and the results are consistent within $1\sigma$.

Nevertheless, there is a systematic trend that $\ob$ and $n_s$
decrease and $\oc$ increases as $\lmin$ grows. Even at $\lmin=100$, 
where the shifts are maximized, they are less than $2\sigma$ in terms
of the new error bars. In terms of the error bars of the model
with $\lmin=2$, the shift is of course larger: for $n_s$
it more than 5$\sigma$, and for $\Omega_\La$ more than 4$\sigma$.
The feature that the error bars on $n_s$ increase more than those
of $\ob$ and $\oc$ may be related to the fact that as
$\lmin$ grows, the pivot scale $k=0.05$ Mpc$^{-1}$ moves
closer to the edge of the data \cite{KurkiSuonio:2004}.

Part of this shift is due to the fact that reionization is neglected.
We know from the absence of the Gunn-Peterson trough in quasar 
spectra that the Universe is reionized at redshifts
$z\lsim 6$, see~\cite{Fan}. The slight decrease towards
smaller scales which is usually attributed to reionization
is now achieved with a somewhat redder spectrum. In order not
to decrease the height of the acoustic peaks, this leads to a
higher value of $\oc$.
A redder spectrum also enhances the amplitude difference
between the well measured first and second peaks. This can be
compensated by a reduction of $\ob$, since a larger $\ob$
means a larger difference between the odd contraction and
even expansion peaks \cite{Durrer:2008}.

However, we have found that reionization is not the dominant effect,
the systematic shift is also present if reionization is included 
in the analysis. We have checked this by including $\tau$ as a 
model parameter. The results of table~\ref{t:linfo} remain valid for
also in this case. The problem is that for $\ell_{\min}\ge 40$ the
value of $\tau$ is degenerate with a renormalization of the amplitude 
(see discussion in Appendix~\ref{ap:reion}) and the best fit value for $\tau$
fluctuates significantly from chain to chain. We therefore prefer to
show the results for $\tau=0$.
Note that the change is larger than the increase
in the error bars. The shape of the one-dimensional probability
distribution for the parameters is not for the most part significantly
distorted, and the two-dimensional distributions do not show strong
changes in the correlation properties as $\lmin$ increases.
Therefore, the error bars do accurately represent
the statistical error even at high $\lmin$.
In other words, the shift in the parameters is systematic,
and is not reflected in the statistical error estimate.

\FIGURE[ht]{
\hspace*{2cm} \epsfig{width=8cm, file=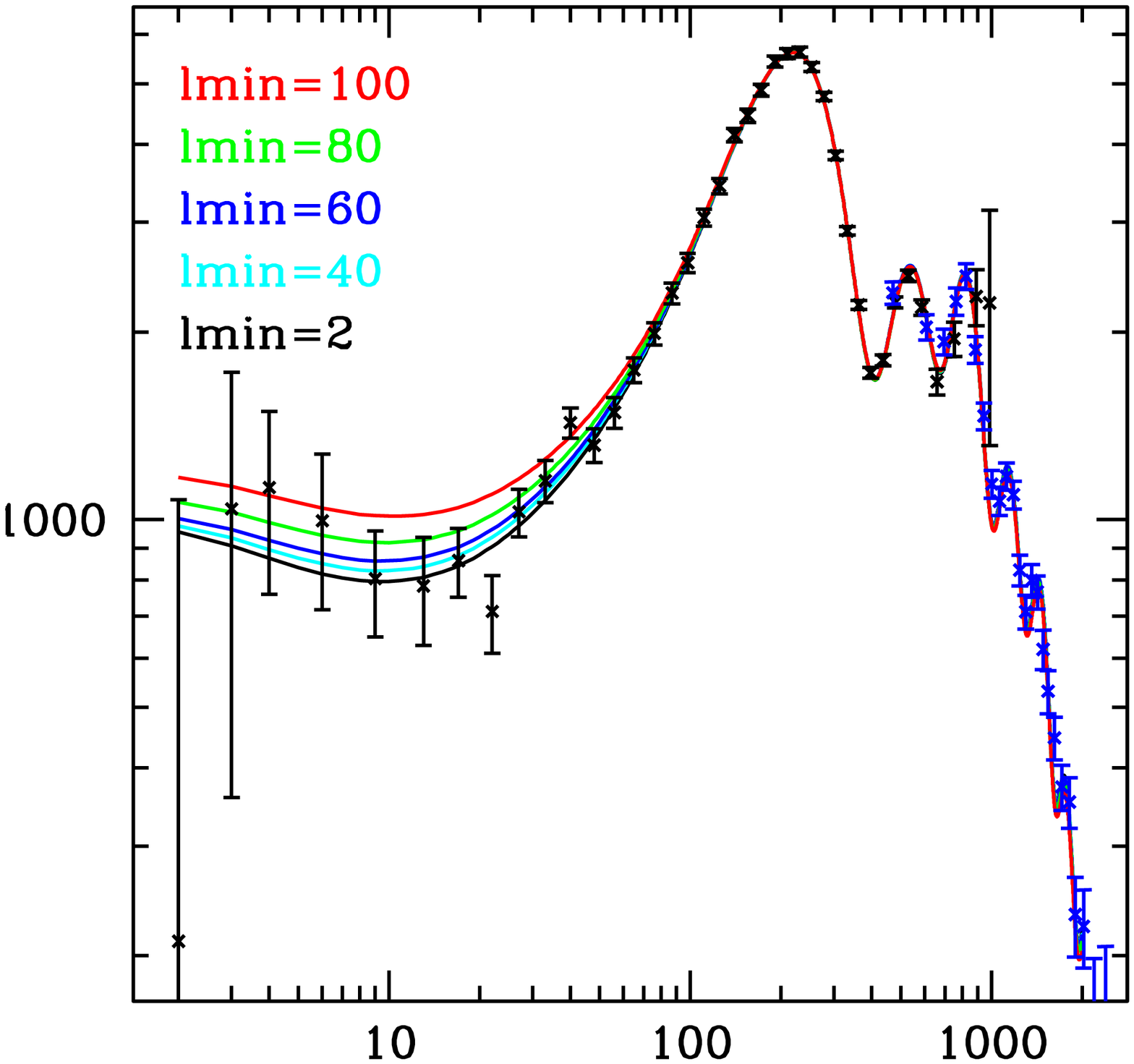} \hspace*{2cm}
\caption{The increase in the large-scale power with increasing
$\lmin$ in the best-fit $\La$CDM models with $\tau=0$. The lowest line
corresponds to a cut at $\ell_{\min}=2$ the subsequent lines have
$\ell_{\min}=40,~60,~80$ and $100$, respectively. At $\ell_{\min}=120$ the 
large scale power no longer increases but it decreases somewhat. The WMAP 
and ACBAR data are superimposed.
The vertical axis is $\ell(\ell+1)C^{TT}/(2\pi)$ in $(\mu K)^2$.
 }\label{f:lcut}}

We conclude that the high $\ell$ data prefer different parameter
values than the data which include the low multipoles.
In \fig{f:lcut} we show the TT power spectra for the
best-fit $\La$CDM models with different $\lmin$.
There is a trend of increasing large-scale power with
higher $\lmin$. In all cases, the overall amplitude is
fixed well by the high $\ell$ data, and the effect is due to
the change in $\ob, \oc$ and $n_s$. We have checked that 
the ISW effect is not the cause: there is a similar
shift for both the $\Lambda$CDM model and the scaled EdS model.
Also, increasing $\lmin$ corresponds to decreasing
$\Omega_\La$ and hence a smaller contribution of the ISW
effect to the low multipoles.

\subsection{Model-independent parameter estimates}
We fix our multipole cut at $\lmin=40$, which roughly
corresponds to neglecting modes which entered the horizon
after $z= 60$. The dependence on the redshift is weak,
$\lmin\propto (1+z)^{1/2}$ for $z\gg1$.
Choosing $z=30$ instead would give $\lmin\approx30$.
The cut at $\lmin=40$ is also motivated by the fact that for
$\ell>40$ reionization is well approximated by a simple rescaling of 
the amplitude, as well as by the multipole dependence of the
late ISW effect, see appendices \ref{ap:scale} and \ref{ap:reion}.

\TABLE[ht]{
%\begin{center}
\begin{tabular}{||l|c|c|c||}
\hline
Parameter & Scaled & $\La$CDM  & $\La$CDM \\
	  & $\lmin=40$ & $\lmin=40$ &$\lmin=2$ \\
          & mean    &  mean   & mean  \\
\hline
$100\ob$ &$2.13\pm 0.05$ & $2.21\pm 0.07$ & $2.24\pm 0.05$ \\
$\oc$ & $0.124\pm 0.007$ & $0.113\pm 0.007$ & $0.111\pm 0.005$ \\
$n_s$ & $0.93\pm 0.02$ & $0.96\pm 0.02$ & $0.97\pm 0.01$ \\
$S$ & $0.91\pm 0.01$ & -- & -- \\
$\Om_\La$ & -- &  $0.72\pm 0.04$ & $0.74\pm 0.03$ \\
$\tau$ & -- & $0.09^{+0.04}_{-0.05}$& $0.09\pm 0.02$ \\
\hline
$\om_m$ & $0.145\pm 0.007$ & $0.136\pm 0.007$ & $0.133\pm 0.005$ \\
$h^{-1} S$ & $2.40\pm 0.03$ &  -- & -- \\
$R$ & $1.77 \pm 0.02$ & $1.73\pm 0.02$ & $1.72\pm 0.02$ \\
$\theta_A$ & $0.593^{\circ}\pm 0.001^{\circ}$ & $0.594^{\circ\, +0.002^\circ}_{\ \, -0.001^{\circ}}$ & $0.593^{\circ} \pm0.002^{\circ}$ \\
$\ell_A$ & $303.7\pm 0.7$ & $303.3\pm 0.8$ & $303.2\pm 0.7$ \\
$D_{A}(z_*)$/Mpc & $12.7\pm0.2$ & $12.9\pm 0.2$ & $13.0\pm 0.1 $ \\
$r_{s}(z_*)$/Mpc & $0.132\pm 0.002$ & $0.134\pm 0.002$ & $0.134\pm 0.001$ \\
$10^{-3}\zeq$ & $3.5\pm0.2$ & $3.3\pm 0.2$ & $3.2\pm 0.1$ \\
$z_*$ & $1094\pm 1$ & $1092\pm 1$ & $1091\pm 1$ \\

\hline
\end{tabular}

%\end{center}
\caption{\label{t:res} The mean values for the scaled model
and the $\La$CDM model. We have used the WMAP5 and ACBAR
data for $\ell\geq\lmin=40$.}
}

In table~\ref{t:res} we give the mean values for our primary parameters
$\ob$, $\oc$, $n_s$ and $S$, as well as some derived parameters.
In addition to the systematic effect discussed above,
this table is our main result. As already 
mentioned, the overall amplitude is treated as a nuisance parameter.
For comparison, we give the corresponding results for the
$\Lambda$CDM model, with non-zero $\tau$.
We use $\lmin=40$ in both cases.
The $\La$CDM values are in good agreement with the WMAP5
results~\cite{Dunkley:2008} and have comparable error bars.
For the scaled model, the errors in $\oc$ and $n_s$ are slightly larger
than those of the $\La$CDM model with $\ell_{\min}=2$.  We  attribute this
to the fact that we start at $\lmin=40$.
Furthermore, our spectral index is somewhat redder, $n_s=0.93$
compared to $n_s=0.96$.
This shift is also clearly seen in the one-dimensional
likelihood functions for the scaled model and the
$\La$CDM model, shown in figure~\ref{f:shift1d}.
However, these parameter changes are within one 
standard deviation and are therefore not statistically significant.
It is impressive how accurately present CMB data determine $\ell_A$.
The relative error is less than 0.3\% for both the scaled model
and $\La$CDM.
The error in the other parameters related to the angular diameter 
distance, $S, h^{-1}S, R$ and $D_A$, as well as $r_s$,
is about 1\%. The errors for $\ob, \oc$ and $n_s$
are less than 3\%, 6\% and 2\%, respectively.

In table~\ref{t:cov} we give the covariance matrix between
the different variables, and in figure~\ref{f:shift2d} we
show selected two-dimensional likelihoods. We see that $R$
and $S$ are strongly positively correlated with 
$\oc$ and $\omm$. In contrast, $D_A$ is strongly
anti-correlated with $\oc$ and $\omm$.
This can be understood by writing $D_A =S D_{A,EdS}$
and noting that $D_{A,EdS}\propto h^{-1}=\omm^{-1/2}$.
The variable  $\ell_A$ is nearly uncorrelated with $\omm$, but it is quite
correlated with $\ob$ and correspondingly also with $n_s$.
Since most of the statistical weight of the WMAP data come from the 
first and second peaks, $n_s$ and $\ob$ are strongly correlated even
if the full WMAP data (with $\lmin=2$) are taken into account
\cite{Komatsu:2008}. This correlation becomes stronger
as some of the low $\ell$ data are omitted.

The standard deviations for the scaled EdS model are
somewhat smaller than those of the $\La$CDM model for
the same $\lmin$.
However, this does not mean that the fit is better,
 only that the well-fitting region is somewhat  smaller.
Error bars for a model can be small simply because different
parts of the data prefer different regions
of parameter space, so that the fit is
good only in some small overlap region. In the present case, 
the scaled model and the $\La$CDM model are
comparably good fits to the data for $\lmin\ge20$.
In table~\ref{t:like} we show $-2\log{\cal L}$, where ${\cal L}$
is the likelihood of the best-fit, as a function of $\lmin$.
There are only differences of $\approx1$ in $-2\log{\cal L}$,
which is the same order as the differences between different
chains of the same model.

%\begin{sidewaystable}
%\centering
%\begin{table}
%\centering
%{\scalebox{0.7}{
\TABLE[htb!]
{
\begin{tabular}{|lrrrrrrrrrr|}
\hline
\hline 
& $\ob$ & $\oc$ & $n_{s}$ & $S$ & $\omm$ & $h^{-1}S$ & $\ell_A$ & $D_A(z_*)$ & $r_s(z_*)$ & $z_*$ \\
 \hline
 \hline
$\scriptsize{\ob}$ & $1.00$ & $-0.31$ & $0.84$ & $-0.49$ & $-0.23$ & $-0.06$ & $-0.56$ & $0.02$ & $0.14$ & $-0.88$
\\
  $\oc$ & $-0.31$ & $1.00$ & $-0.51$ & $0.96$ & $1.00$ & $-0.91$ & $0.05$ & $-0.94$ & $-0.98$ & $0.72$ \\
  $n_s$ & $0.84$ & $-0.51$ & $1.00$ & $-0.65$ & $-0.45$ & $0.19$ & $-0.51$ & $0.26$ & $0.38$ & $-0.86$ \\
  $S$ & $-0.49$ & $0.96$ & $-0.65$ & $1.00$ & $0.94$ & $-0.78$ & $0.30$ & $-0.82$ & $-0.91$ & $0.83$ \\
  $\omm$ & $-0.23$ & $1.00$ & $-0.45$ & $0.94$ & $1.00$ & $-0.94$ & $-0.004$ & $-0.96$ & $-1.00$ & $0.66$ \\
  $h^{-1}S$ & $-0.06$ & $-0.91$ & $0.19$ & $-0.78$ & $-0.94$ & $1.00$ & $0.32$ & $1.00$ & $0.96$ & $-0.40$ \\
  $\ell_A$ & $-0.56$ & $0.05$ & $-0.51$ & $0.30$ & $-0.004$ & $0.32$ & $1.00$ & $0.27$ & $0.06$ & $0.43$ \\
  $D_A(z_*)$ & $0.02$ & $-0.94$ & $0.26$ & $-0.82$ & $-0.96$ & $1.00$ & $0.27$ & $1.00$ & $0.98$ & $-0.47$ \\
  $r_s(z_*)$ & $0.14$ & $-0.98$ & $0.38$ & $-0.91$ & $-1.00$ & $0.96$ & $0.06$ & $0.98$ & $1.00$ & $-0.58$ \\
  $z_*$ & $-0.88$ & $0.72$ & $-0.86$ & $0.83$ & $0.66$ & $-0.40$ & $0.43$ & $-0.47$ & $-0.58$ & $1.00$ \\   
\hline
\end{tabular}
%}
\caption{\label{t:cov} The normalized covariance matrix for the scaled 
model. We have used the WMAP5 and ACBAR data for $\ell\ge\lmin=40$.
At this level of precision, the correlation coefficients
of $R$ are the same as those of $S$, and those of $\theta_A$
are minus those of $\ell_A$.}
}
%\end{table} 

\FIGURE[ht]{
\includegraphics[width=12.5cm]{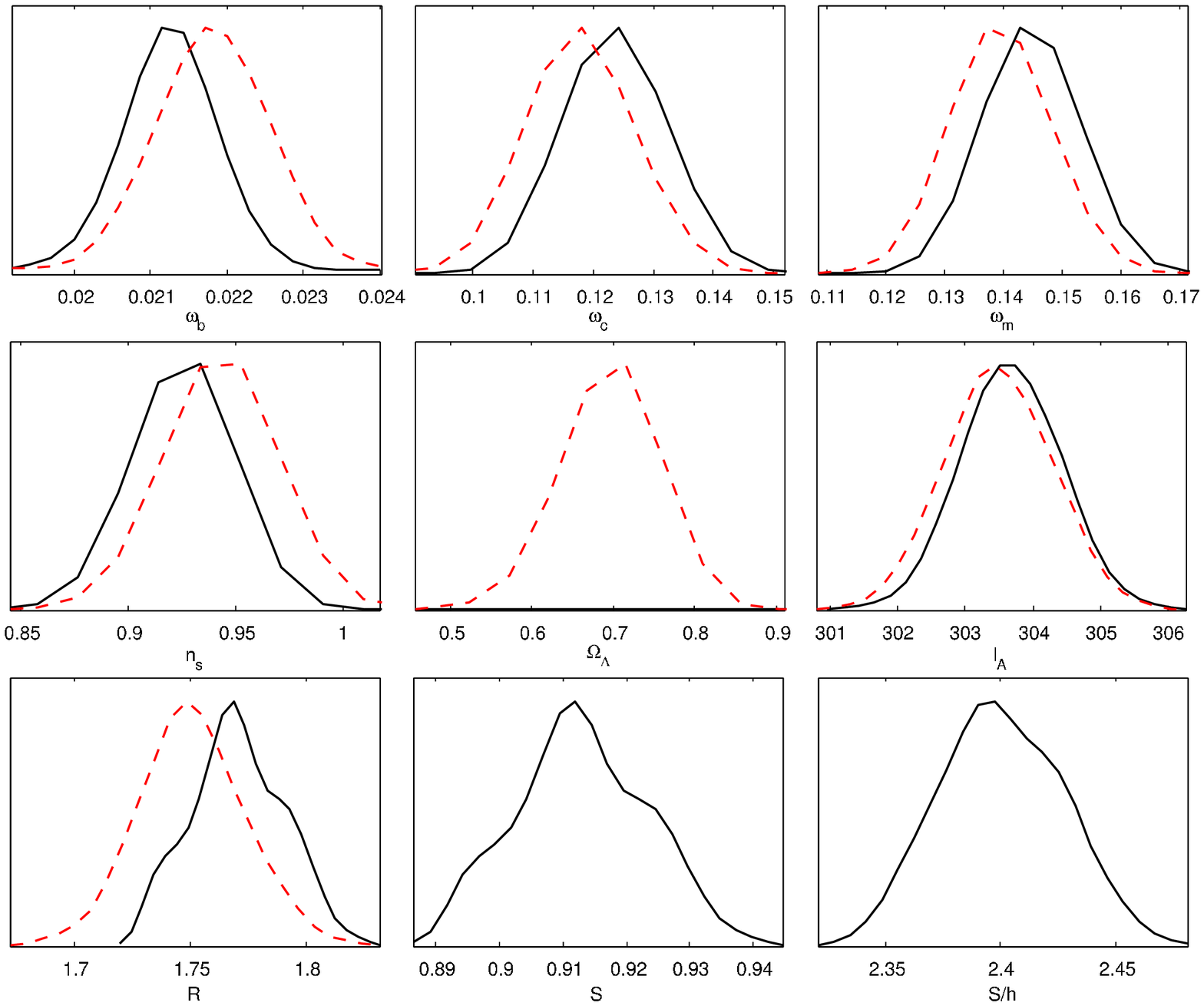}
\caption{\label{f:shift1d} One-dimensional likelihoods for
the scaled model (black, solid) and the $\La$CDM model (red, dashed).
We have used the WMAP5 and ACBAR data for $\ell\ge\lmin=40$.}
}
\clearpage
\FIGURE[ht]{
\includegraphics[width=13cm]{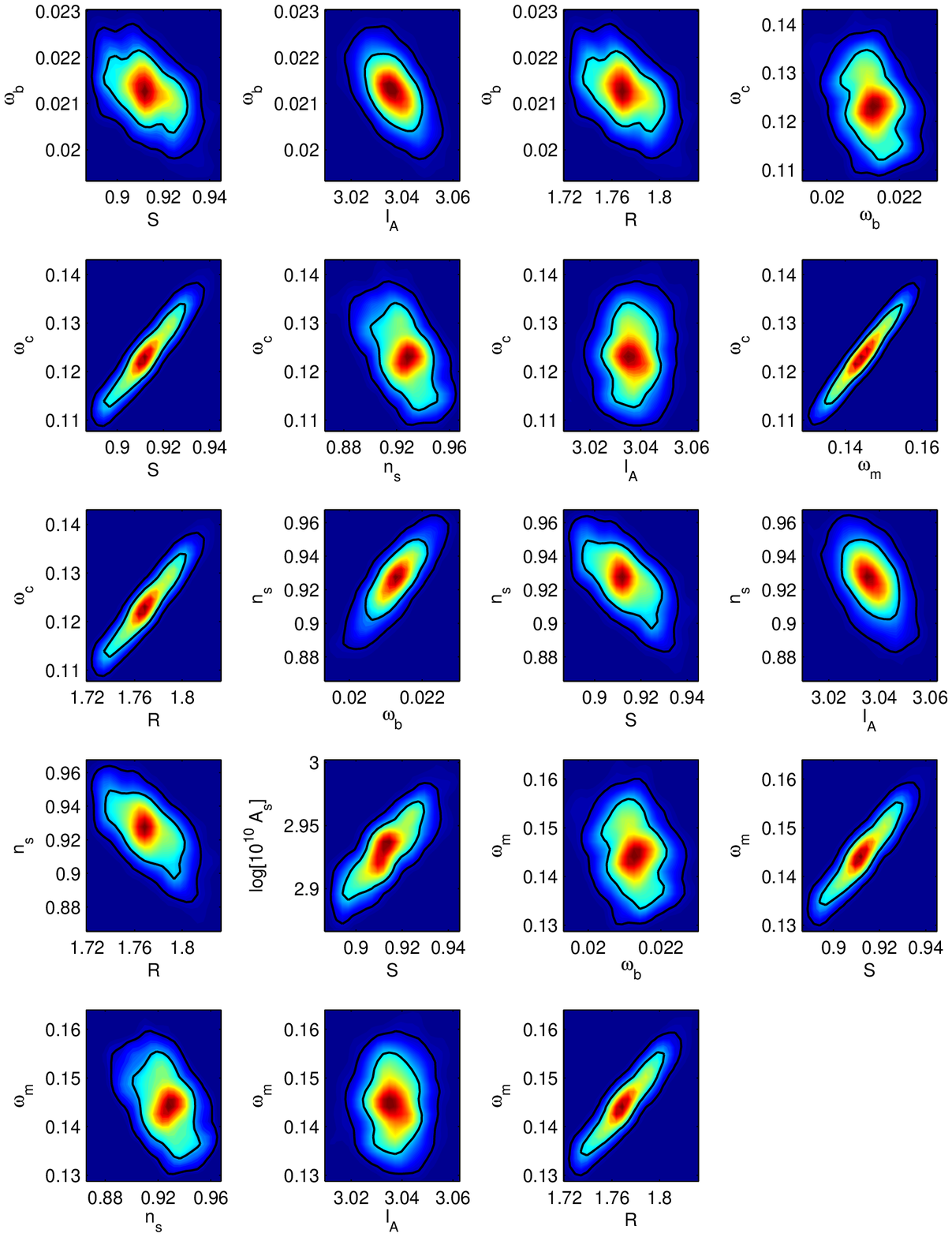} %2Dfinal.ps}
\caption{\label{f:shift2d} Two-dimensional likelihoods for the scaled 
model. We have used the WMAP5 and ACBAR data for $\ell\ge\lmin=40$.}
}

\TABLE[ht]{
\begin{tabular}{||l|c|c|c|c||}
\hline
$\lmin$ & nr. of points & $-2$log$\mathcal{L}$ &  $-2$log$\mathcal{L}$ &
 $-2$log$\mathcal{L}$ \\
  & $N(\lmin)$  &   scaled &  standard, $\tau=0$ &standard, $\tau\neq0$ \\
\hline
 2 & 2591 & 2717.12 & 2715.78 & 2695.29 \\
20 & 1385 & 1508.20& 1507.41 & 1507.72\\
40 & 1345 & 1382.52 & 1381.24 & 1381.24\\
60 & 1305 & 1234.44 & 1233.23& 1233.46\\
80 & 1265 & 1073.03 & 1072.01 & 1072.16\\
\hline
\end{tabular}
\caption {The log of the likelihood $\mathcal{L}$ as function of $\lmin$.
In the second column we give the number of $C_\ell$ estimates
(including the polarization data) except for the case $\lmin=2$
where a pixel-likelihood is added. For $\lmin\geq20$,
$N(\lmin)=994+427-2(\lmin-1)$,
which is the number of multipoles for the TT
(WMAP5 and ACBAR data) and TE (WMAP5 data) spectra minus
twice the number of cut multipoles. The only significant difference
between models appears in the first row with $\lmin=2$,
where the $\Lambda$CDM model with $\tau\neq 0$ is clearly favoured. 
\label{t:like}}
}

\subsection{Discussion}

The CMB contains information about the
distance to the last scattering surface, the baryon density, the
matter density and the primordial power spectrum (here
taken to be a power law), which can be extracted
independently of the model used to describe the late universe.
In particular, the angular diameter distance to the last scattering
surface is a factor of $S=0.91\pm 0.01$ smaller than in an EdS
Universe with the same mean matter density,
$\omm=0.145\pm 0.006$. With baryon density $\ob=0.0213\pm0.001$
and spectral index $n_s=0.93\pm0.03$, an EdS model scaled by this factor
is a good fit to the present CMB data, apart from the low multipoles.
Of course such a model is in complete disagreement with
local measurements of the Hubble parameter and supernova
observations. If we want to agree with the local
value $H_0=$ (60--70) km/s/Mpc, the observed distance
is instead longer than in an EdS model
by the factor $h^{-1}S\approx$ 1.4--1.7.
From the CMB we cannot determine at which point between
last scattering and today the distance evolution diverges
from the EdS case; from supernova observations, we
know that this happens between a redshift of order unity and today.
Any viable cosmological model has to explain this change in
the distance scale, whether the reason is dark energy, modified
gravity or large deviations from the FL geometry.

Constraints on $R$, $\ell_A$ and other parameters have been
 presented earlier in
\cite{Komatsu:2008, Elgaroy:2007, Wang:2007, Corasaniti:2007, Mukherjee:2008, Komatsu:2010},
where the data have been analysed in the context
of different models for dark energy, also taking into account
effects like neutrino masses which we do not consider.
Our mean value for $R$ is larger (and $\ob$ and $n_s$ are
smaller) than in those studies, because of the systematic
shift due to cutting away the low multipoles.
The increase in the error bars is smaller than the change
in the mean values, as they do not take into account the
systematic shift. The shift indicates that different parts
of the data prefer different parameter values, which frustrates
the effort to give precise model-independent error bars,
because the only way to reduce model-dependence is to exclude
the part of the data which is most likely subject to unknown
physical effects. We think that cutting the multipoles below
$\lmin=40$ strikes a good balance between reducing
model-dependence and not discarding data needlessly.

The cosmological parameter most robustly determined
by the CMB in a model-independent manner is the ratio
$\ell_A=\pi D_A(z_*)/r_s(z_*)$, which does not undergo
a systematic shift with increasing $\lmin$, unlike
$\ob, \oc, n_s, R$ or $D_A(z_*)$.
It is interesting that as low multipoles are
cut, the spectral index becomes smaller, making the
evidence for violation of scale-invariance in the
initial conditions stronger. For $\lmin\ge80$, values
$n_s<0.9$ are within 1$\sigma$ of the mean.
As for the baryon density, the shift towards smaller
values is well within the constraint
$1.9\leq100\ob\leq2.4$ (95\% C.L.) from
Big Bang Nucleosynthesis \cite{Amsler:2008}.
Our value for $\oc$ is more than 2$\sigma$ away from the
$\La$CDM value with no multipole cut, while the
error bars increase only by 26\%.
This model-dependence suggests caution about the
 value and the error bars of $\oc$
which enter into codes such as DarkSUSY \cite{DSUSY}.

In order to be independent of late-time cosmology, we cannot
take into account low $\ell$ results for the CMB anisotropies.
In the final  parameters quoted in
table~\ref{t:res} we have used the data for $\ell\ge \lmin=40$.
At first sight one might hope that our analysis could be
significantly improved once the Planck data with precise
$C_\ell$'s up to $\ell\approx2500$ will be available.
However, for $\ell\gtrsim1000$ CMB lensing can no longer be
neglected for data with a precision better than about
4\%  for the anisotropies and 10\% for the polarisation
\cite{Lewis:2006, Durrer:2008}. But
 lensing and other second order effects depend on the details
of the late-time cosmology. Hence our model-independent analysis
has to be restricted to the interval of roughly $40\le \ell\le 800$.
Higher $\ell$ data can only be used if the error bars are
sufficiently large. For ACBAR this is still marginally possible, but
with Planck systematic errors due to late-time effects will have to be 
added to the high $\ell$ data.
Increased precision in the multipole range $40\le \ell\le 800$
also has to be balanced against contamination by model-dependent
secondary effects. We therefore do not expect a substantial
improvement of our results from future data.

\section{Conclusion}

We have analysed the CMB data in a way which is
independent of the details of late-time cosmology,
i.e. the cosmology at redshifts $z\lsim 60$. The results
we have obtained are therefore valid for most models
of late-time cosmology, whether they include dark energy,
modified gravity, a local void or backreaction.

We have presented model-independent limits on $\ob$, $\oc$, $n_s$ and the
angular diameter distance to the last scattering surface $D_A(z_*)$,
or its ratio with the sound horizon at last scattering,
$\theta_A=r_s(z_*)/D_A(z_*)$.
The present CMB data give an extraordinarily precise 
measurement of $\theta_A$, which every realistic model
of the late universe must agree with. We can 
summarize the final result by
\bea\label{e:fin}
  100\ob &=& 2.13 \pm 0.05 \ , \qquad \oc = 0.124 \pm 0.007 \el
  n_s &=& 0.93 \pm 0.02 \ , \qquad \theta_A = 0.593^\circ\pm 0.001^\circ  \ .
\eea

\noindent Note that the values of $\oc$ and $\ob$
actually determine the matter and baryon density at last scattering
via the relation $\rho_x(z_*) = (1+z_*)^3(H_0/h)^2\om_x$. The values
of the densities today may be different e.g. if dark
matter decays at late times \cite{DMdecay}.

In summary, every model which satisfies equations \re{e:fin}
will automatically be in agreement with the present CMB data for
$\ell\geq40$. Only lower $\ell$ CMB data, large scale structure, lensing
and other observations can distinguish between models which have the
above values for $\ob, \oc, n_s$ and $\theta_A$. 

We have also found that there is a systematic shift in the
cosmological parameters as more low $\ell$ data are cut.
As more data from low multipoles is removed, $\ob$ and $n_s$
decrease, while $\oc$ becomes larger. These changes keep
the power spectrum at small scales fixed, but tend to increase the
amplitude on large scales. These changes are not
reflected in the statistical error bars: 
the small angle data prefer different parameter values
than the full set of CMB data. This trend is visible to
at least $\lmin=100$.
Whether this behaviour has any connection with the various
directional features at low 
multipoles~\cite{anom, Hansen:2008, Francis:2009, Bennett:2010},
is not clear.

\acknowledgments

We thank Domenico Sapone who participated in the beginning
of this project. This work is supported by the Swiss National
Science Foundation.

\appendix

\section{The scale parameter approximation} \label{ap:scale}

In this appendix we illustrate the accuracy of the
scale parameter approximation for the high multipoles.
We consider spectra for FL models with non-zero spatial
curvature or cosmological constant, compared with the
Einstein-de Sitter result scaled with the parameter $S$ as discussed
in \sec{sec:method}. We keep the matter densities fixed to
the WMAP5 best-fit values $\ob=0.023$ and $\oc=0.11$
\cite{Dunkley:2008}. Neglecting the  contribution of
radiation, the scale parameter in these models is
\be
  S \simeq \frac{\sqrt{\om_m}}{2}\int_0^{z_*} 
 \frac{dz}{\sqrt{\om_m(1+z)^3+\om_K(1+z)^2 +h^2-\om_m-\om_K}} \ ,
\ee
where $\om_K\equiv\Om_Kh^2$ and $h^2-\om_m-\om_K =\Om_\La h^2 \equiv 
\om_\Lambda$.

In \fig{f:curvatureTT} we show the TT spectrum for models
with positive or negative spatial curvature and the scaled
model. The spectra lie on top of each other for
$\ell\gtrsim 20$, except for large negative spatial curvature.
In \fig{f:curvatureTE} and \fig{f:curvatureEE} we show the TE and EE spectra.
The scaled curves are practically indistinguishable from the
exact ones at all multipoles, even for large negative spatial curvature.
In \fig{f:lambda}, we show the spectra for models
with positive cosmological constant compared with the
scaled model. The scaling approximation is excellent for
all of the spectra for $\ell\gtrsim20$.

\FIGURE[ht]{
\includegraphics[width=13cm]{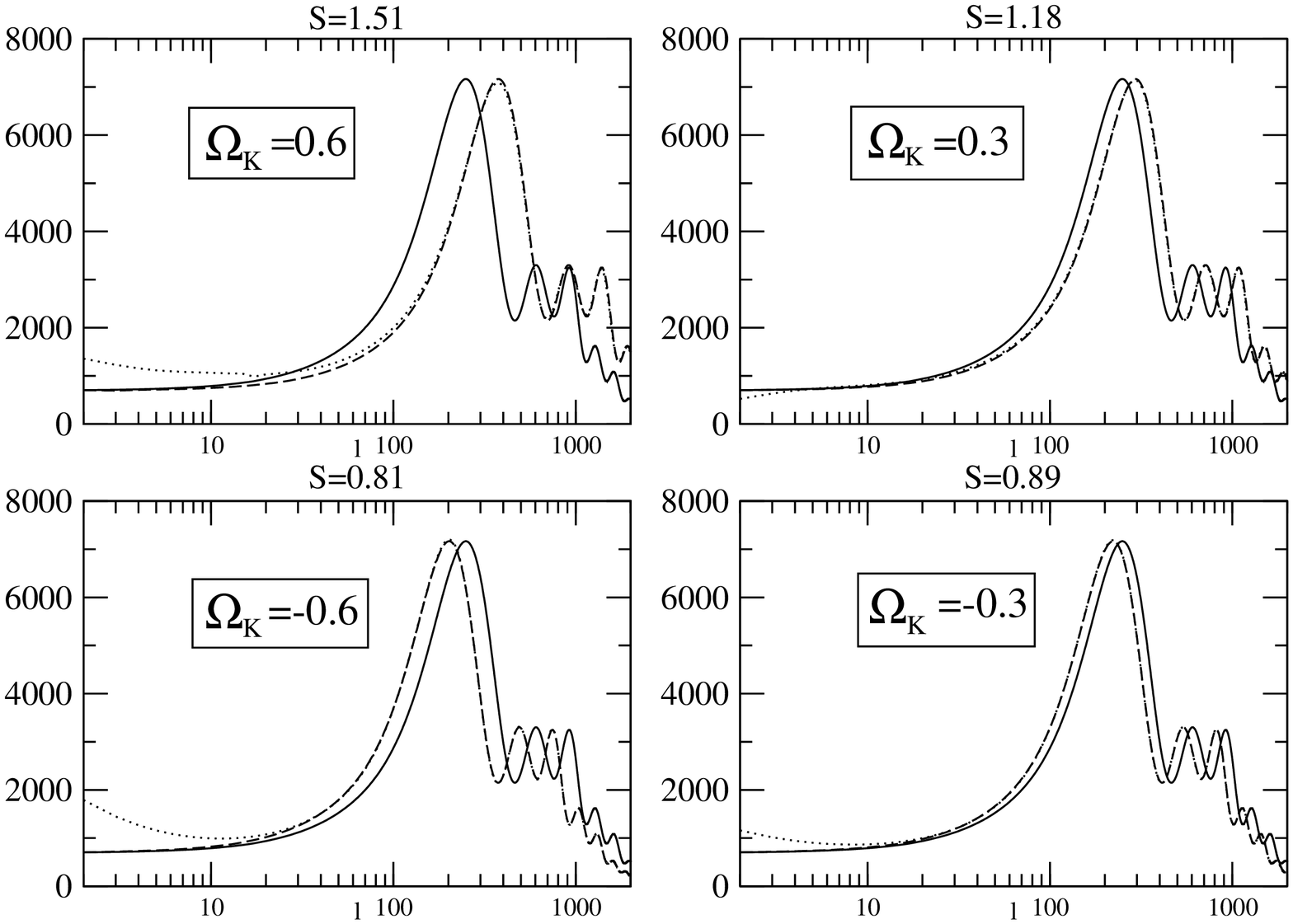} \vspace*{1cm}
\caption{The TT spectra for models with $\Omega_\La=0, \Omega_K\neq0$.
The solid curve corresponds to the Einstein-de Sitter universe,
the dotted curve corresponds to a model with $\Omega_K$ as specified
in the panels, and the dashed curve shows the Einstein-de Sitter universe
power spectrum scaled with $S$.
The vertical axis is $\ell(\ell+1)C^{TT}_\ell /(2\pi)$ in $(\mu K)^2$.
\label{f:curvatureTT}} }
\clearpage

\FIGURE[ht]{
\includegraphics[width=13cm]{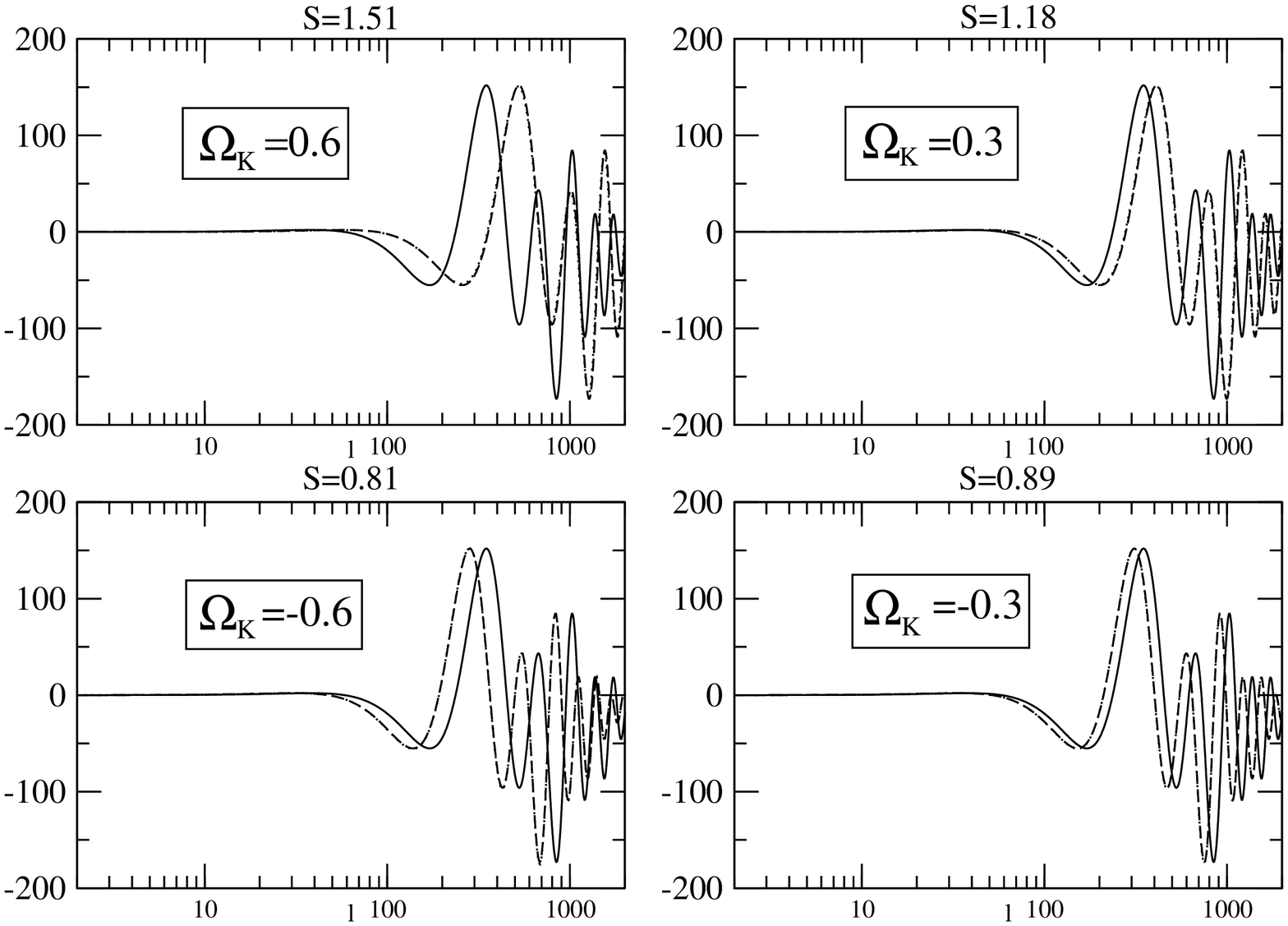} \vspace*{1cm}
\caption{As in figure~\ref{f:curvatureTT}, but for the TE spectra.
The dotted curves are invisible since they are completely overlaid by 
the dashed ones (scaled model).   
\label{f:curvatureTE}}}[ht]

\FIGURE[ht]{
\includegraphics[width=13cm]{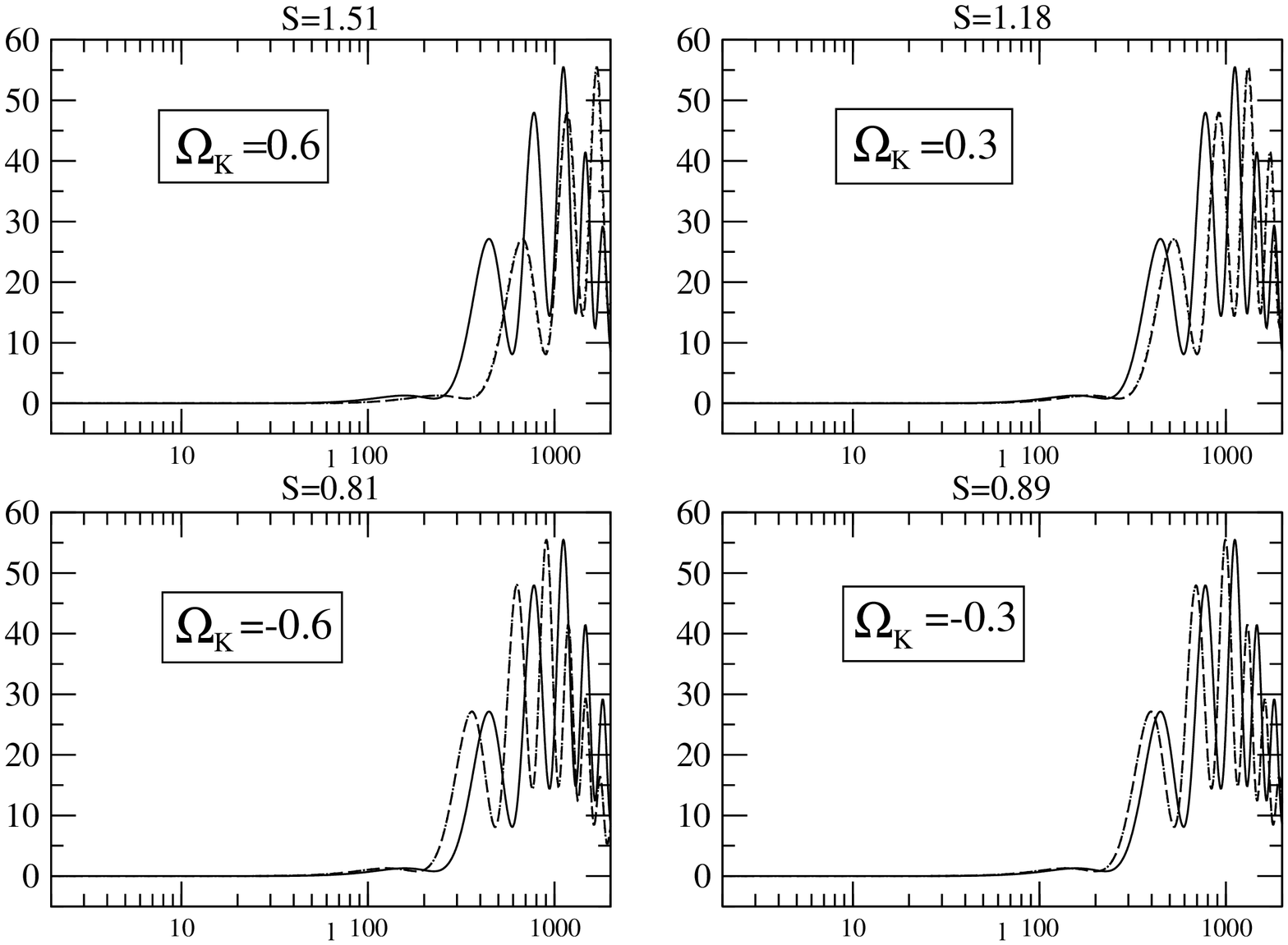}\vspace*{1cm} 
\caption{As in figure~\ref{f:curvatureTT}, but for the EE spectra.  
The dotted curves are invisible since they are completely overlaid by 
the dashed ones (scaled model).
\label{f:curvatureEE}}}\clearpage

\FIGURE[ht]{
\epsfig{width=13cm, file=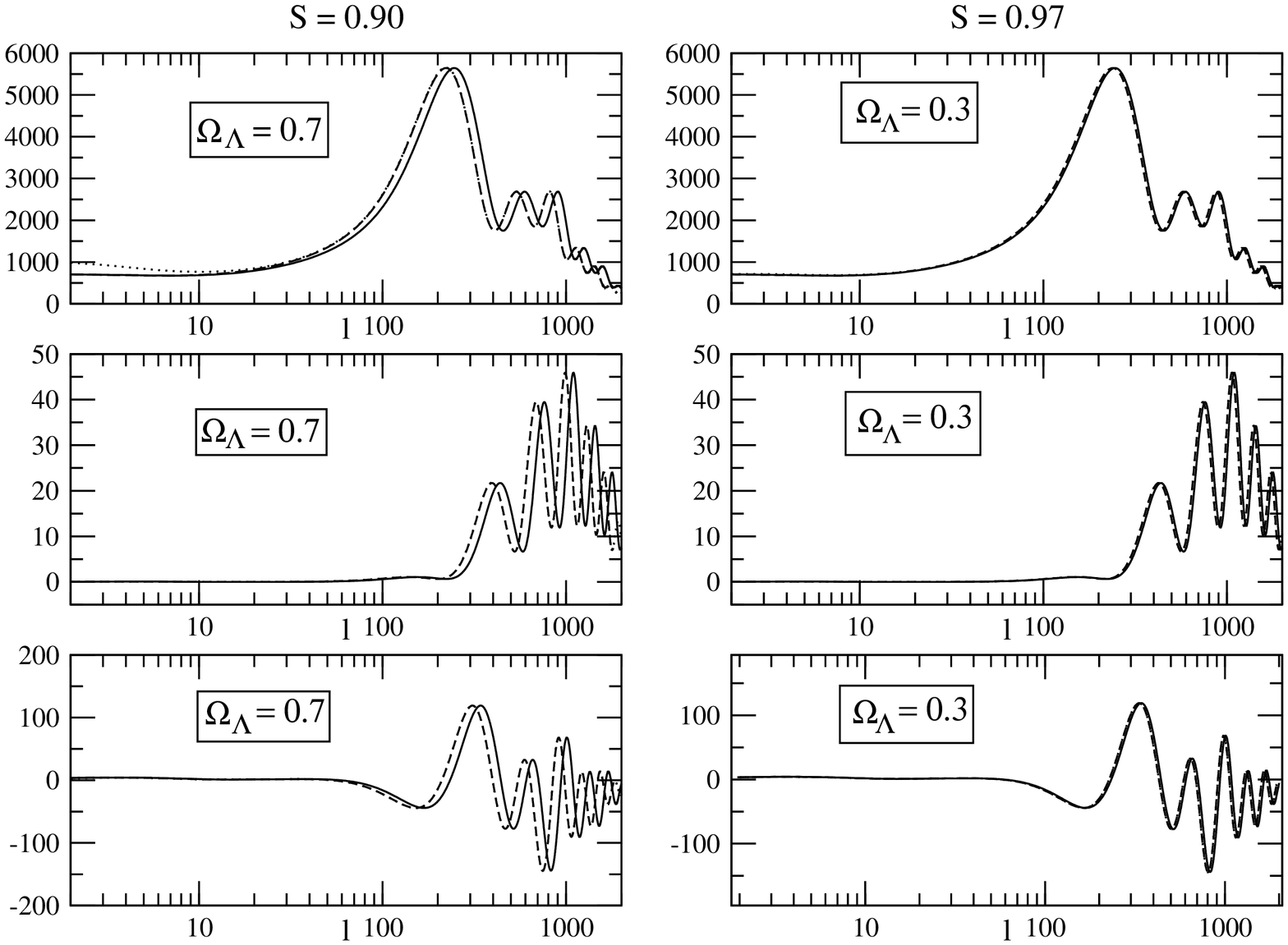}\vspace*{1cm}
\caption{As in \fig{f:curvatureTT}, but for $\Omega_\La\neq0, \Omega_K=0$.
We consider two different values for $\Omega_{\Lambda}$,
corresponding to the two columns. The rows from top to bottom
are the TT, EE and TE spectra.
\label{f:lambda}}}

\section{Reionization} \label{ap:reion}

In this appendix we study the effect of reionization on the angular 
power spectrum of the CMB. If the baryons are reionized at redshift $\zre$,
the effect on scales which are of the order of the
horizon size at the time is complicated, and leads to additional
polarization and a scale-dependent reduction of the amplitude
of anisotropies. However, on scales which are well inside the horizon,
the rescattering of photons simply reduces the amplitude of CMB temperature
and polarization anisotropies by roughly the same amount on all scales.
This effect can therefore be absorbed in a renormalization of the spectrum.
In figure~\ref{f:reTT} we show the TT spectrum with and without reionization
for the best-fit $\La$CDM model, as well as the relative
difference of the spectrum with and without reionization.
For $\ell\ge 40 $, renormalizing the spectrum with a 
constant reproduces the effect of reionization within about 1.5\%. We 
have done the same with the temperature--polarization cross-correlation 
and the polarization spectra. Also there renormalization is a very good 
approximation (better than 0.5\% on average) for $\ell\ge 40$,
see figures~\ref{f:reET} and~\ref{f:reEE}. To obtain the spectra with 
$\tau=0.1$, we have multiplied the spectra with $\tau=0$ by the factor 
$0.82$.

\FIGURE[ht]{
\epsfig{width=14cm,file=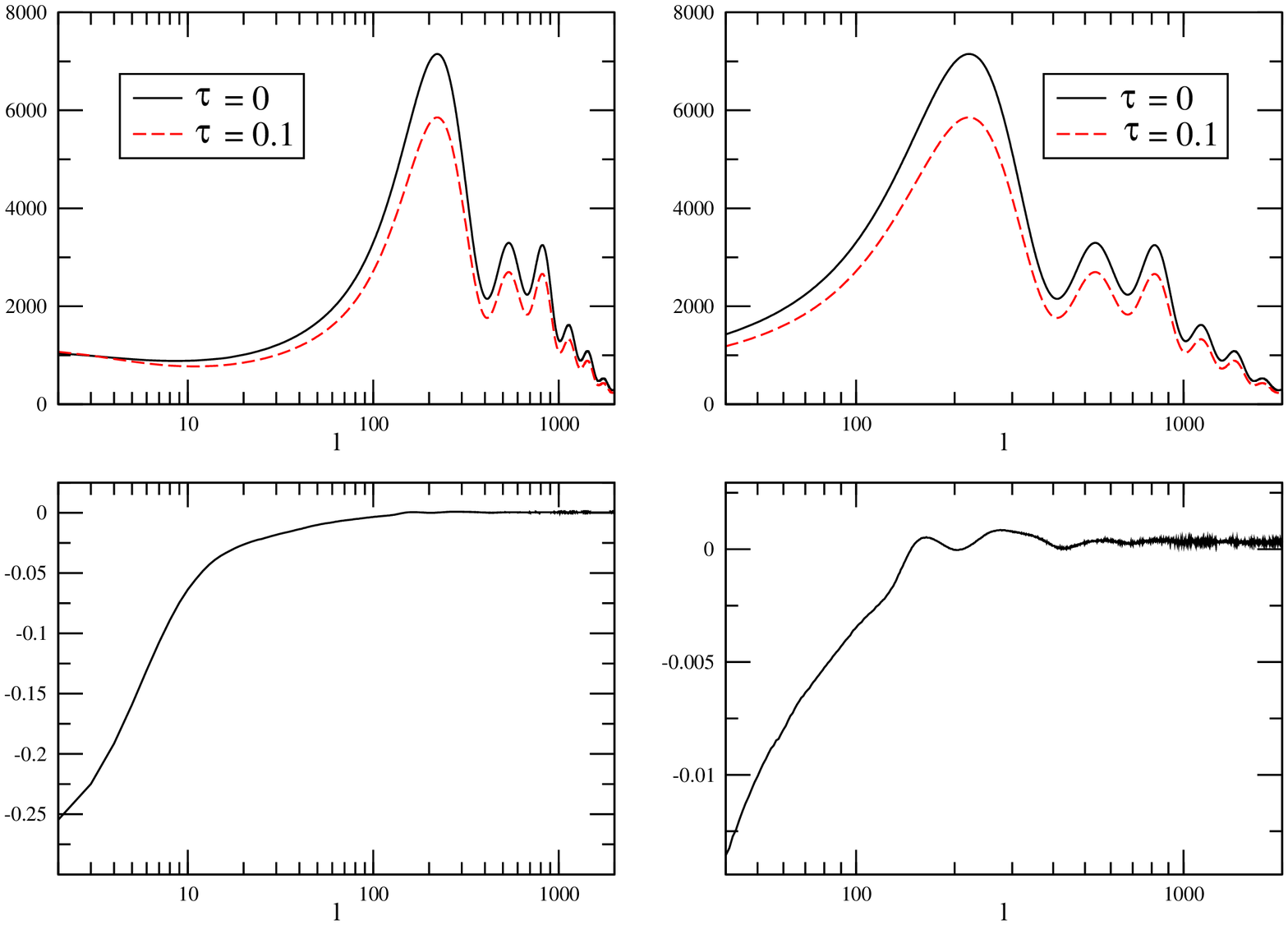} \vspace*{0.4cm}\\
\caption{The TT power spectrum with (dashed, red) and 
without (solid, black) reionization for
optical depth $\tau=0.1$ for $\ell\ge 2$ (left upper panel) and $\ell\ge 40$ 
(right upper panel). For the upper panels,
the vertical axis is $\ell(\ell+1)C^{TT}/(2\pi)$ in $(\mu K)^2$.
In the lower panel we show the relative difference
between the spectrum with and without reionization, when the latter
is simply rescaled by a constant.
For low $\ell$'s, the differences are substantial, up to 25\%, 
but for the values $\ell\ge40$ we consider, the difference 
is less than 2\%.
\label{f:reTT}}
} %\end{figure}

\FIGURE[ht]{ 
\epsfig{width=14cm,file=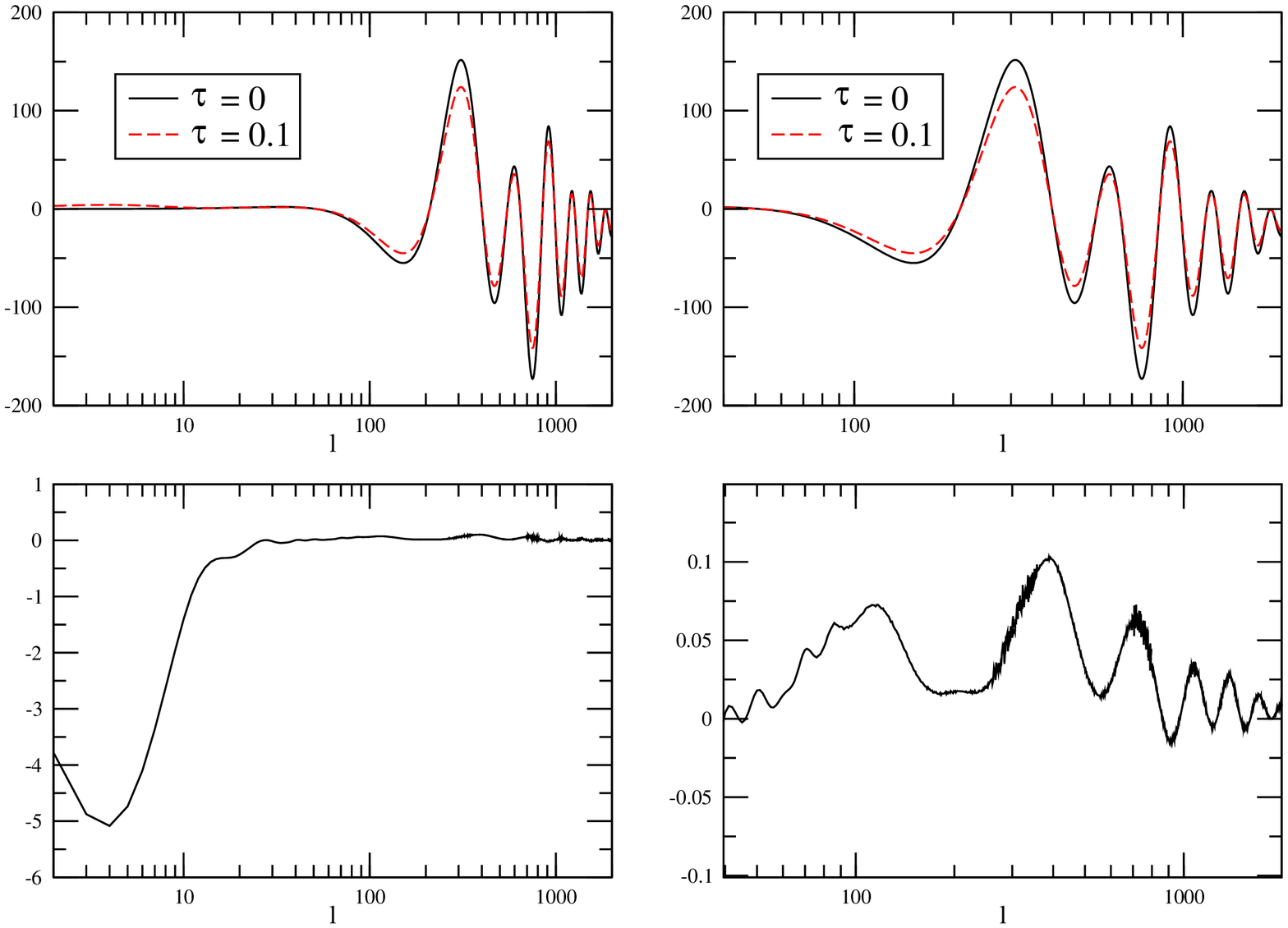}  \vspace*{0.4cm} \\
%\vspace{0.5cm}
\caption{The TE correlation spectrum with (dashed, red) and 
without (solid, black) reionization for optical depth $\tau=0.1$
for $\ell\ge 2$ (left upper panel) and $\ell\ge 40$ (right upper panel). 
The vertical axis is $\ell(\ell+1)C^{TT}/(2\pi)$ in $(\mu K)^2$. In the lower 
panel we show the difference between the spectrum with and without
reionization, when the latter is simply rescaled by a constant.
For the values $\ell\ge40$ we consider, the difference is below
$0.1 (\mu K)^2$.
\label{f:reET}}
}

\FIGURE[ht]{ 
\epsfig{width=14cm,file=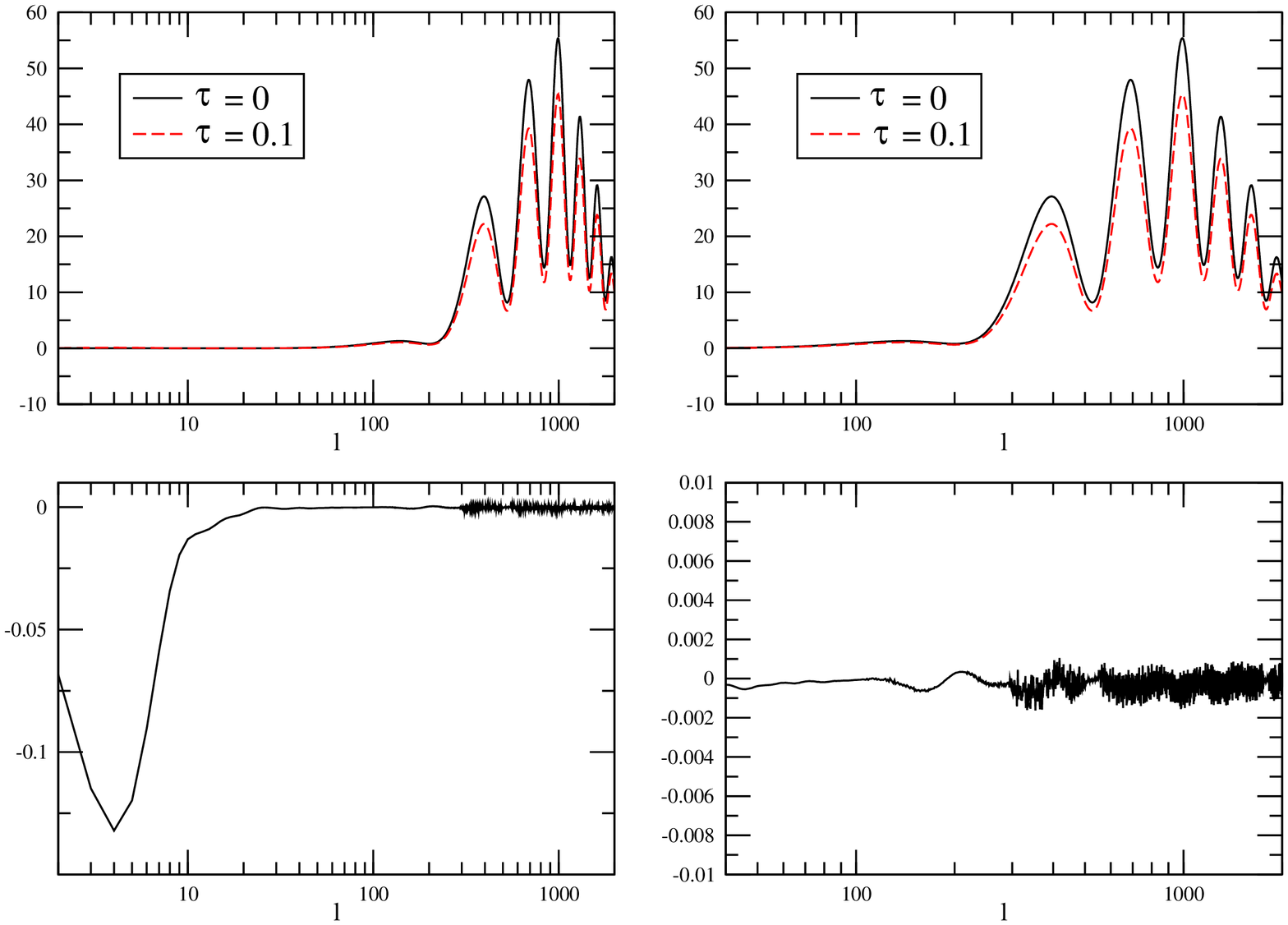}  \vspace*{0.4cm}\\
\caption{As in figure~\ref{f:reET}, but for the EE power spectrum.
For $\ell\ge40$, the difference is below
$0.002 (\mu K)^2$.
\label{f:reEE}}
}

\end{document}